\documentclass[preprint,12pt]{elsarticle}

\usepackage{amssymb}

\usepackage{lineno}
\journal{Journal of Nuclear Instrument and Methods A}

\biboptions{sort&compress}

\begin{document}

\begin{frontmatter}

\title{Characterization of two SiPM arrays from Hamamatsu and Onsemi for liquid argon detector}

\renewcommand{\thefootnote}{\fnsymbol{footnote}}
\author{
T.A.~Wang$^{a}$, 
C.~Guo$^{b,c,d,}$,
X.H.~Liang$^{e}$,
L.~Wang$^{b,c}\footnote{Corresponding author.  Tel:~+86-18273488483. E-mail address: leiwang@ihep.ac.cn (L.~Wang). }$,
M.Y.~Guan$^{b,c,d,}$,  
C.G.~Yang$^{a,b,d}$, 
J.C.~Liu$^{a,b,d}$,
F.Y.~Lin$^{a,b,d}\footnote{Corresponding author. Tel:~+86-13756090065. E-mail address: linfengyuan\_0116@163.com (F.Y.~Lin).}$,
}


\address{
	${^a}${State Key Laboratory of High Power Semiconductor Laser, College of Physics, Changchun University of Science and Technology, Changchun, Jilin, China}
	
	${^b}${Experimental Physics Division, Institute of High Energy Physics, Chinese Academy of Sciences, Beijing, China}
	
	${^c}${School of Physics, University of Chinese Academy of Sciences, Beijing, China}
	
	${^d}${State Key Laboratory of Particle Detection and Electronics, Beijing, China}
	
	${^e}${Astro-particle Physics Division, Institute of High Energy Physics, Chinese Academy of Science, Beijing, China}
}


\begin{abstract}

	Silicon photomultiplier (SiPM), a new type of photosensor, is considered a substitute for traditional photomultiplier tube (PMT) in the next generation of dark matter and neutrino detectors, especially in noble gas detectors like liquid argon. However, the design of compact SiPM arrays and their cryogenic electronics that can work in liquid argon is barely developed. Thus, two candidate SiPM arrays from Hamamatsu and Onsemi were selected to verify the feasibility and effectiveness of the design. In this work, we successfully developed a cryogenic electronics read-out system that connects and works with 1-inch 4$\times$4 SiPM arrays at 87~K. The power dissipation of amplifiers is less than 10 $\mu$W/mm$^2$. Furthermore, multiply significant characteristics of both types of SiPM arrays were measured at liquid argon temperature, such as dark count rate (DCR), breakdown voltage (V${_{bd}}$), single photoelectron (SPE) performance, signal to noise ratio (SNR) and correlated signal probability. 
\end{abstract}

\begin{keyword}
	SiPM array, Liquid argon, Cryogenic electronics, Dark count rate, Correlated signals. 
\end{keyword}

\end{frontmatter}


\section{Introduction}\label{sec:section1}

Silicon photomultiplier (SiPM) was initially developed in Russia in the mid-1980s~\cite{SiPM-his}. After decades of development, it offers an attractive alternative to photomultiplier tubes (PMT) in noble gas detectors. In the United States, the Deep Underground Neutrino Experiment (DUNE) employed SiPMs in a large liquid argon time projection chamber(LArTPC)~\cite{C-DUNE} and reported characterizations of SiPMs developed for the DUNE Photon Detection System~\cite{SiPM-DUNE}. DarkSide-20k was designed as a 20-tonne fiducial mass dual-phase LArTPC with SiPM-based cryogenic photosensors~\cite{DarkSide-20k}. Kazutaka Aoyama of Waseda University in Japan is also trying to apply SiPM arrays for liquid argon detector~\cite{SiPM-Japan}. In contrast to traditional PMTs, SiPM arrays have many advantages~\cite{SiPM-applications,OnsemiSiPM}, such as the excellent uniformity of the single photoelectron response, lower operating voltage ($<$100~V), lower radioactive background, and lower prices.

Therefore, a type of SiPM array that can work in LAr is crucial for next-generation LAr detectors. Our previous work demonstrated the feasibility of the single-chip S13370-6050CN SiPM at this temperature~\cite{VUV4_SiPM}. However, the ceramic package of the single-chip SiPM makes a 100\% compact photon-sensitive area of scintillator detectors impossible. Thus, constructing the SiPM array without any package edge is necessary.

Consequently, two types of SiPM arrays from Hamamatsu (S14161) and Onsemi (J-60035) were selected to address this issue. So far, their characterizations have been barely studied at liquid argon temperature. In this paper, multiple characterizations of both SiPM arrays were measured, which include changes in breakdown voltage(V${_{bd}}$), SPE performance, DCR, SNR, and correlated noise with different temperatures from 237~K down to 87~K. This work's primary purpose is to study the possibility of applying the two types of SiPM in liquid argon detectors.

\section{Experimental setup}\label{sec:section2}

\begin{figure}[htbp]
	\centering
	\includegraphics[width=11cm]{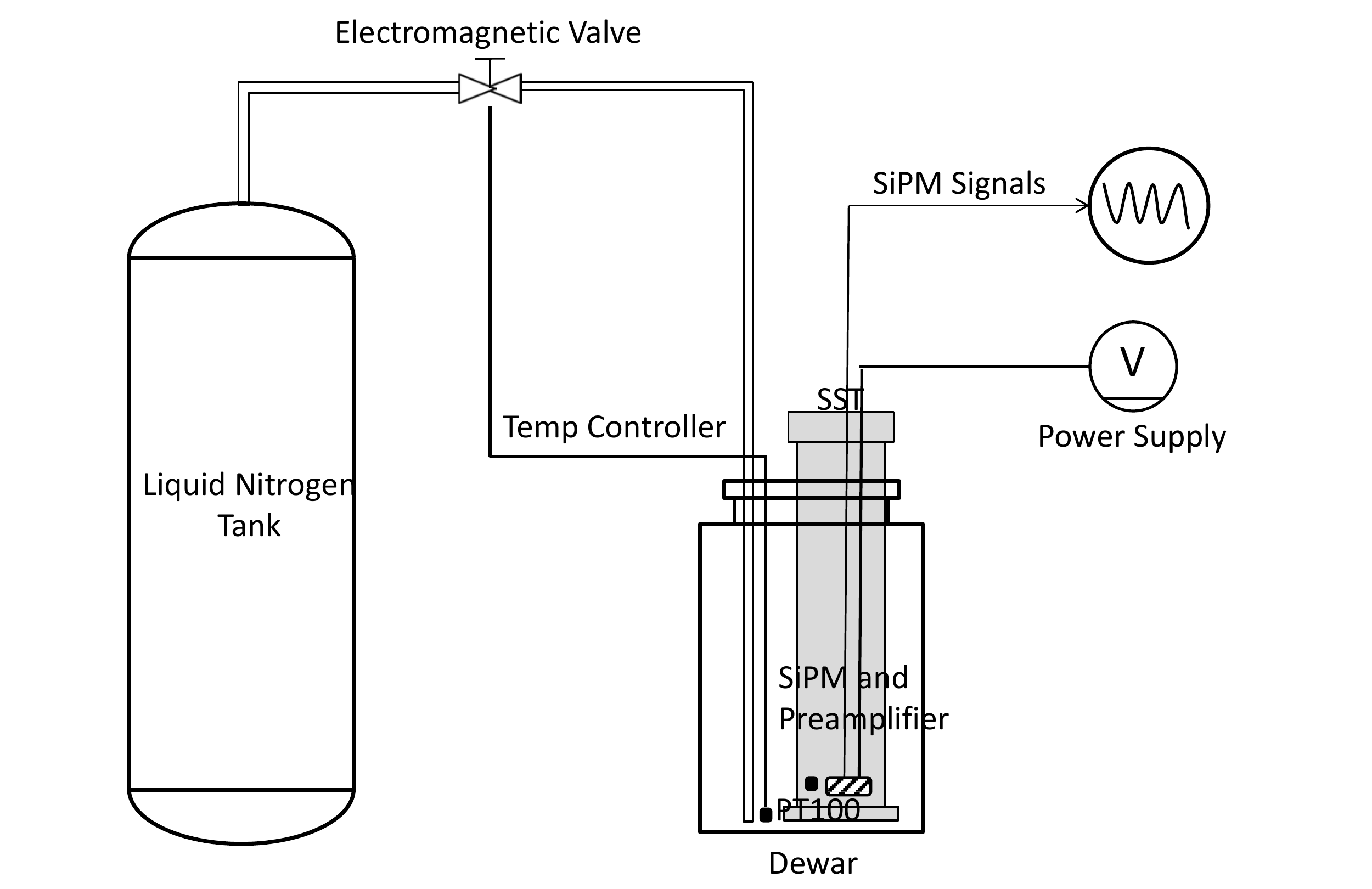}
	\caption{\label{setup_graph} Schematic of the experimental setup.}
\end{figure} 

The schematic diagram of the cryogenic system used in the measurement is shown in Fig.~$\ref{setup_graph}$, which is composed of a liquid nitrogen tank, a dewar tank, a stainless steel (SST) chamber, two temperature sensors, and an electromagnetic valve. Two SiPM arrays and their cryogenic electronics could be cooled down to liquid argon temperature (87~K) in one hour simultaneously. More details of the experimental setup can be found in Ref.~\cite{VUV4_SiPM}. Before cooling down, the SST would be filled with nitrogen to avoid the influence of water vapor condensation on the SiPM surface.

\section{Electronics} \label{sec:section3}


\begin{figure}[htbp]
	\centering
	\includegraphics[width=9cm]{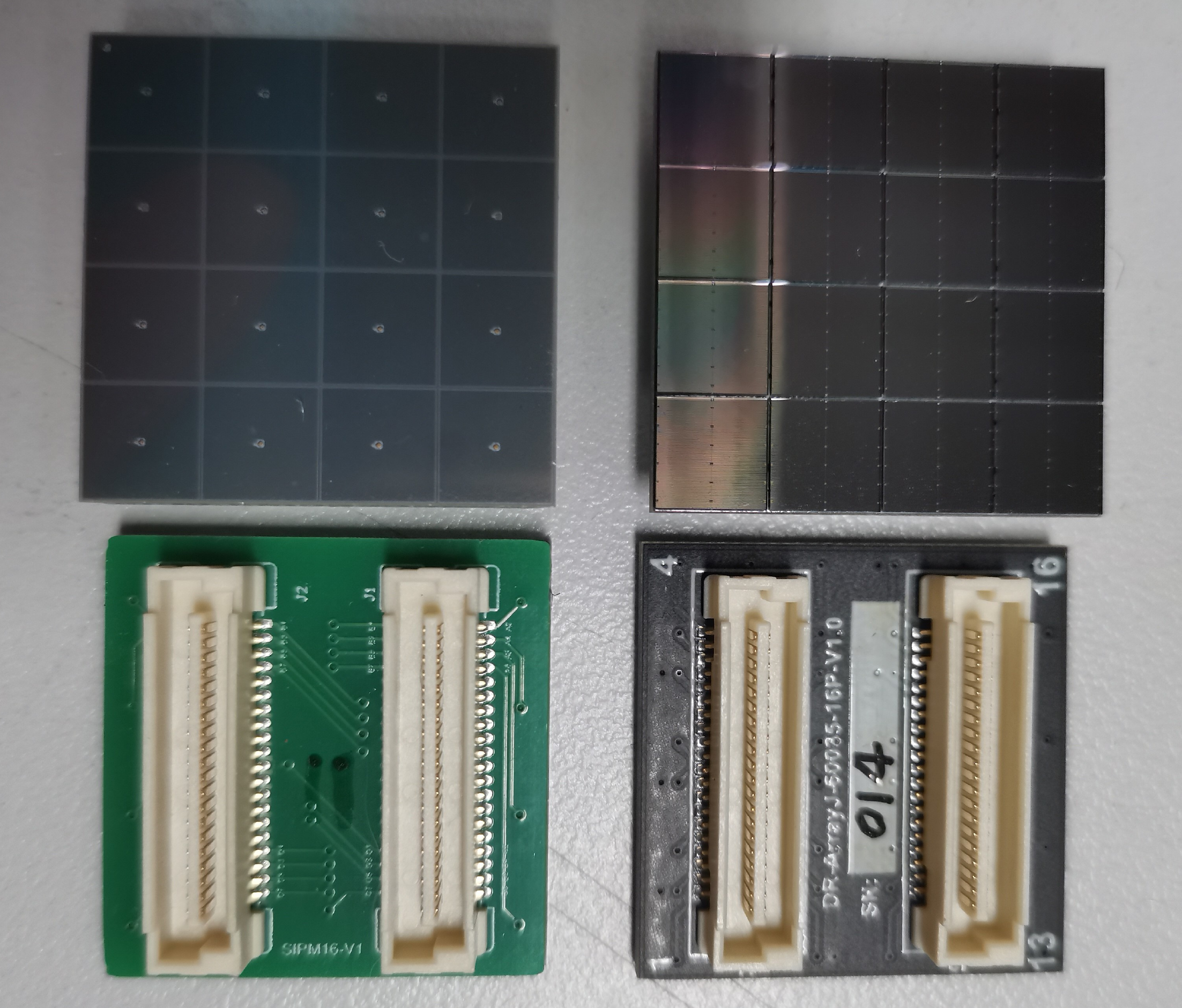}
	\caption{\label{SiPM_arrays_figure} Left: a Hamamatsu S14161-6050HS SiPM array. Right: an Onsemi J-Series 60035 SiPM array.} 	
\end{figure}

In the measurement, a 4$\times$4 Hamamatsu S14161-6050HS SiPM array and a 4$\times$4 Onsemi J-Series 60035 SiPM array were tested. In Fig.~\ref{SiPM_arrays_figure}, the left side shows the 4$\times$4 Hamamatsu S14161-6050HS SiPM array~\cite{HamaDS} and the right side shows the 4$\times$4 Onsemi J-Series 60035 SiPM array~\cite{OnsemiDS}. They are both 25$\times$25~mm$^{2}$ in size, and both consist of isolated sixteen single-chip SiPMs on the top layer. On the back layer, a board-to-board connector was designed for plug-and-play conveniently.

\subsection{Pre-amplifying circuit and printed circuit board(PCB)}

\begin{figure}[htbp]
	\centering
	\includegraphics[width=9cm]{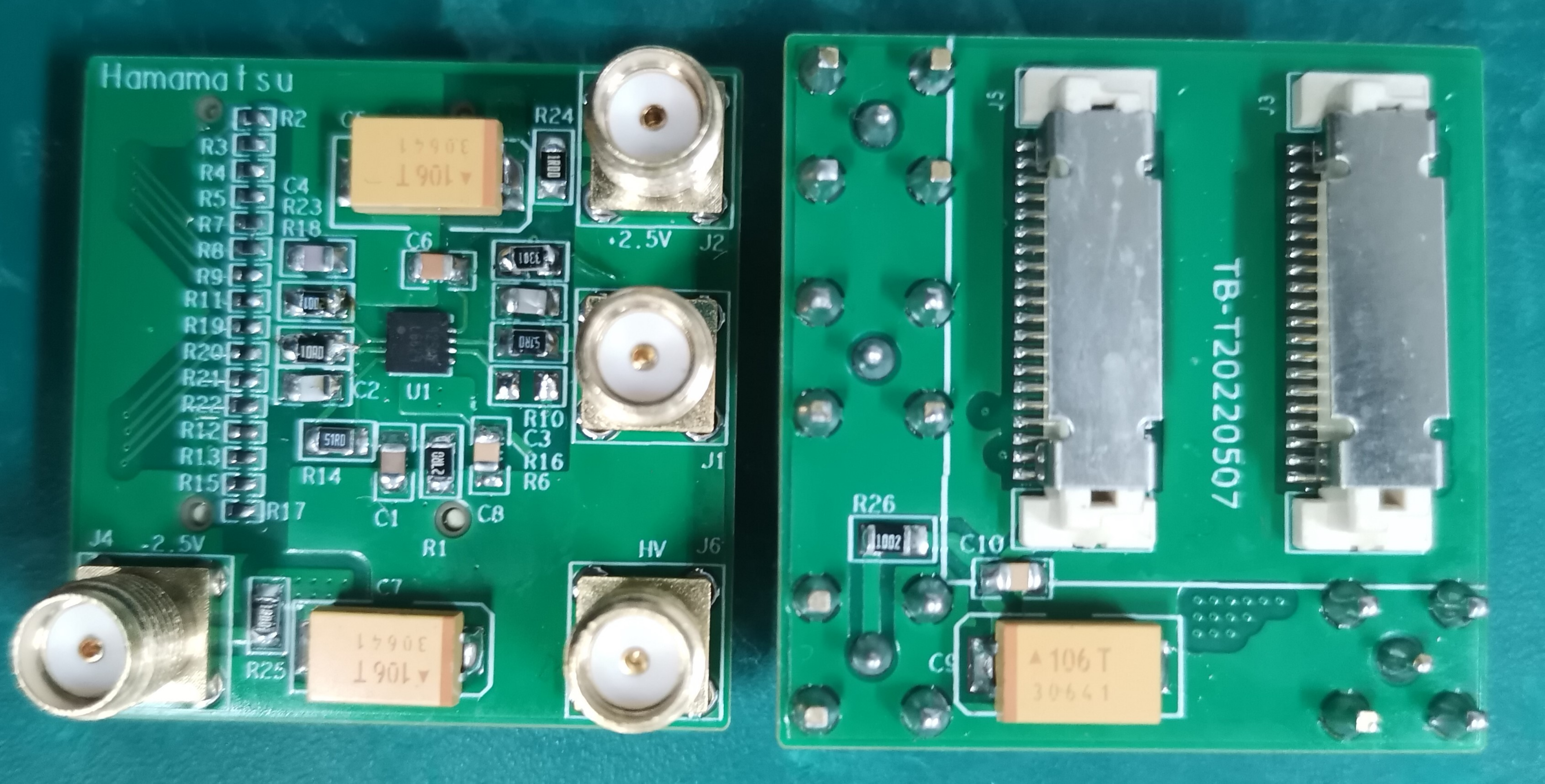}
	\caption{\label{PCBoard_figure} The PCB of the readout circuit.} 	
\end{figure}

The PCB for pre-amplifying signals from SiPM arrays is shown in Fig.~$\ref{PCBoard_figure}$, with the size of 33$\times$33~mm$^{2}$. Four SMA connectors (one for voltage loading, one for SiPM signal output, and two for pre-amplifier power supplies) were adopted for the external connection interface of the circuit. They are widely used small thread coaxial connectors with excellent performance, such as a wide frequency band and high reliability. For detecting weak photocurrent signals from the SiPM output, a trans-impedance amplifier (TIA) is required. LMH6629 is one of a series of transimpedance amplifiers of Texas Instruments, with a high speed and ultra-low noise~\cite{LMH6629DS,TIA_con}. Our previous work has proved its stability at liquid argon temperature~\cite{VUV4_SiPM,SiPMLAr}.

\begin{figure}[htbp]
	\centering
	\includegraphics[width=14cm]{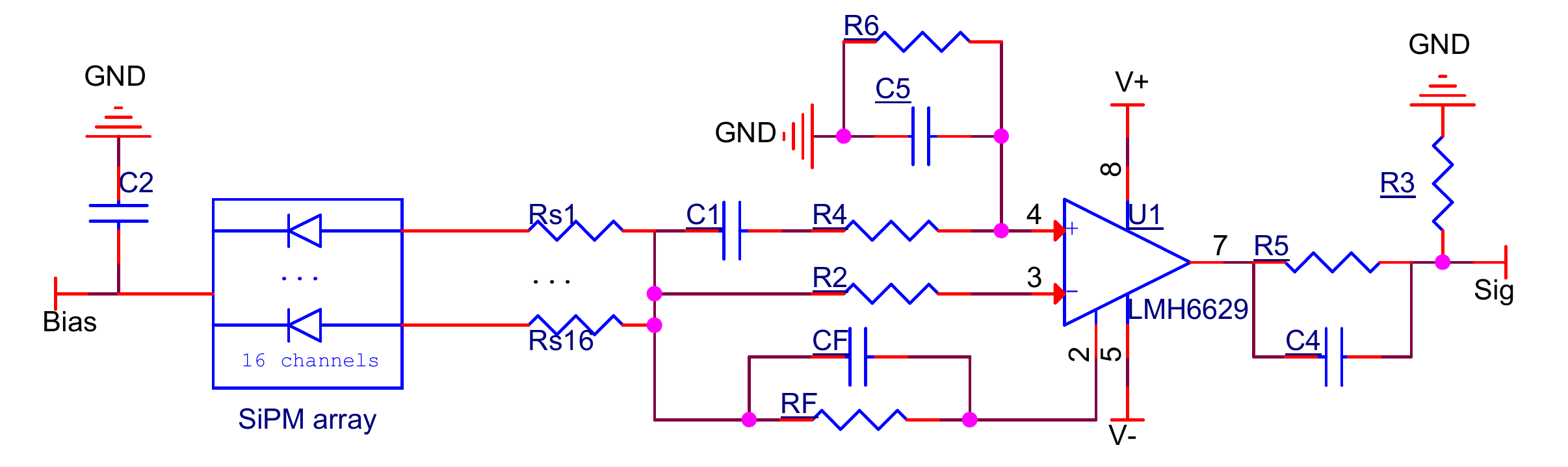}
	\caption{\label{circuit_figure} Circuit diagram of the readout.} 	
\end{figure}

The circuit schematic diagram is shown in Fig.~$\ref{circuit_figure}$. The circuit adopted the typical trans-impedance connection method to obtain the best signal-to-noise ratio. For instance, the resistors R4 and R2 at the inverting and non-inverting input were required to avoid high-frequency oscillations under the cryogenic temperature~\cite{Preamplifier}. The grounding resistance R6 at the non-inverting input was used as a compensation resistance, ensuring the symmetry of the external resistance between the input terminals of the amplifier. Moreover, the capacitor C5 can minimize the additional noise of R6. The integral capacitor C${_F}$ formed a low-pass filter circuit with R${_F}$ to eliminate the influence of the high frequency. It could also protect the circuit from self-excitation. Sixteen channels of the SiPM array were connected to the amplifier circuit in parallel.

LMH6629 converts I$_{in}$, the photocurrent signals of SiPM arrays, into amplified voltage signals V$_{out}$. The trans-impedance gain is determined by the value of R$_{F}$~\cite{TIA_con} as shown in Eq.~\ref{Eq.V_out}. 
\begin{equation}
V_{out}=I_{in}\times R_F
\label{Eq.V_out}
\end{equation}
\begin{equation}
\frac{\int V_{out}dt}{R_{load}}=\frac{\int I_{in}dt\times R_F}{R_{load}}
\label{Eq.Q}
\end{equation}

In Eq.~\ref{Eq.Q}, R${_{load}}$ is the matched resistance of the oscilloscope that is used for data acquisition. $\int I_{in}dt$ is the SiPM output charge, and $\frac{\int V_{out}dt}{R_{load}}$ at the left side of the equation is the charge recorded by the data acquisition. Thus, the gain for amplifying the SiPM output charge could be defined finally as G${_{amp}}$=R${_{F}}$/R${_{load}}$, as shown in Eq.~\ref{Eq.G_SiPM}.
\begin{equation}
\frac{\int V_{out}dt}{R_{load}}=G_{amp}\times \int I_{in}dt
\label{Eq.G_SiPM}
\end{equation}

If the left side of Eq.~\ref{Eq.G_SiPM} is replaced by the charge of the single photoelectron Q${_{SPE}}$, and G${_{SiPM}}$ is defined as the gain of SiPM, Eq.~\ref{Eq.G_SiPM} could be modified to: 
\begin{equation}
	Q_{SPE}=G_{amp}\times G_{SiPM} \times Q_{e}
	\label{Eq.Amp}
\end{equation}

Where Q$_{e}$=1.6$\times$10$^{-16}$~C is the charge of an electron. G$_{SiPM}$ of two types of SiPM arrays at room temperature could be found in their datasheet. However, this parameter is not clear at liquid argon temperature.  According to Eq.~\ref{Eq.Amp}, G$_{SiPM}$ of two SiPM arrays under 87~K were estimated.  Tab.~\ref{RT_LAr} shows their G$_{SiPM}$ separately at room and liquid argon temperatures.

\begin{table}[hbp]
\centering

\begin{tabular}{cccccc|}
	 \hline
	      &            & Onsemi & Hamamatsu \\ \hline
 	      & V${_{bd}}$ & 24.2~V & 38~V \\
	   RT & V${_{over}}$ & 2.5~V & 2.7~V \\
	      & G${_{SiPM}}$ & 2.9$\times$10$^{6}$ & 2.5$\times$10$^{6}$ \\ \hline
	      & V${_{bd}}$ & 20.88~V & 31.3~V \\
	   87~K& V${_{over}}$ & 4~V & 4.2~V \\
	      & G${_{SiPM}}$ & 1.93$\times$10$^{6}$ & 2.16$\times$10$^{6}$ \\ \hline 
\end{tabular}
	\caption{\label{RT_LAr}G${_{SiPM}}$ and V${_{bd}}$ of two SiPM arrays at room temperature (RT) and LAr temperature. G${_{SiPM}}$ is calculated when V${_{over}}$~$\approx$ 4~V.
}
\end{table}

\subsection{DAQ setups}

Both SiPM arrays were placed in the SST chamber together for testing, and two DH 1765-4 DC power supplies provided the operating voltage for them. And two RIGOL DP831A DC power supplies were used for the power supply for two LMH6629. A LeCroy 104Xs-A oscilloscope was used for data acquisition.

\begin{figure}[htbp]
	\centering
	\includegraphics[width=11cm]{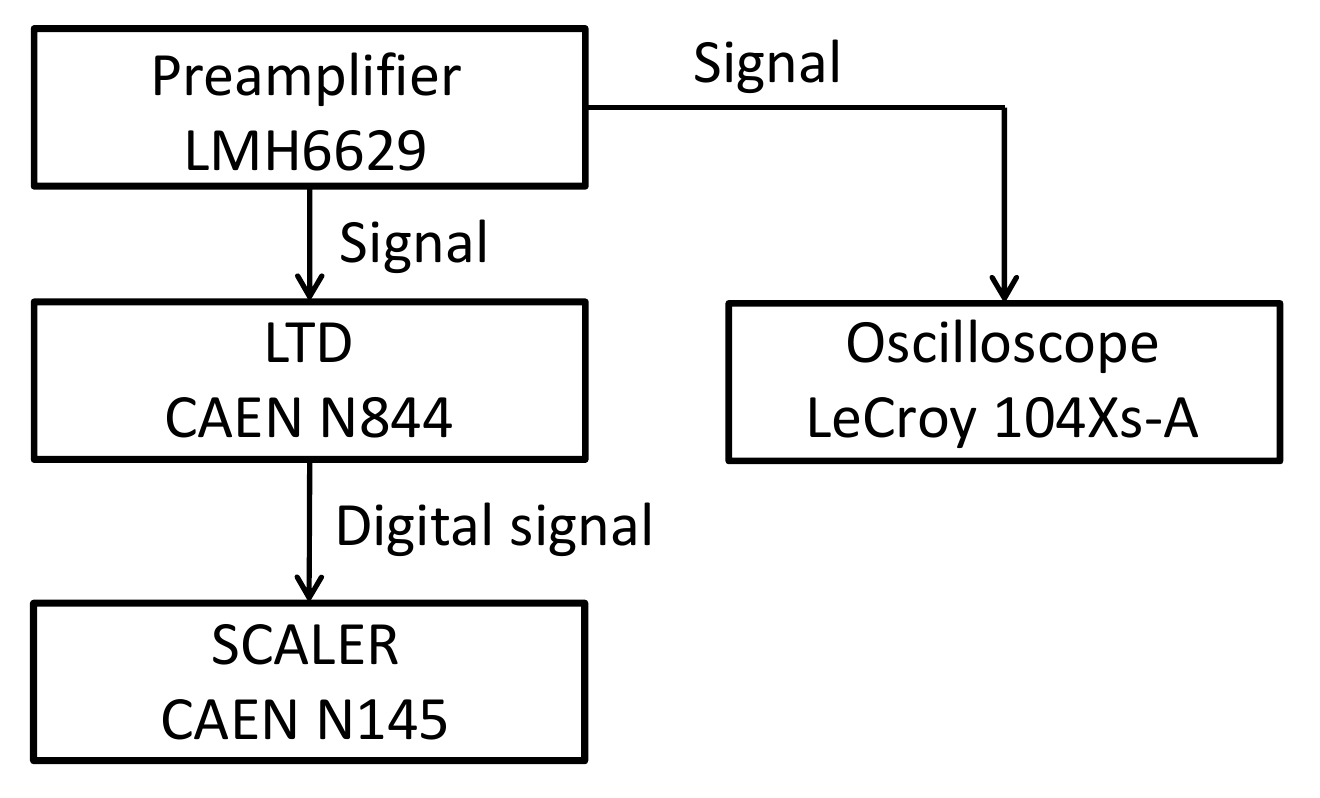}
	\caption{\label{DCR_sch} Schematic of characterizations measurement.} 	
\end{figure}

The measurement scheme is shown in Fig.~$\ref{DCR_sch}$. Three different kinds of DAQ setups have been used in the measurement.

(A)While measuring the breakdown voltage (V$_{bd}$) and signal-to-noise ratio (SNR), the oscilloscope is in self-trigger mode, and the threshold is set to half of the single photoelectron amplitude. The time window is set to 5000 ns. And the data sampling rate of the oscilloscope is set to 1 GS/s.

(B)While measuring the correlated signal probabilities, the setup is the same as the measurement of V$_{bd}$. Except that the oscilloscope time window is set to 20~$\mu$s, the data sampling rate of the oscilloscope is set to 500~MS/s.

(C)While measuring the dark counting rate (DCR), the SiPM array signals are sent to LTD (CAEN, N844). The digital signal output by LTD is forwarded to a scaler (CAEN, N145) for counting. And the threshold is set to half of the single photoelectron amplitude.

\section{Data analysis}\label{sec:section4}

\subsection{Breakdown voltage}

The breakdown voltage (V${_{bd}}$) is the bias point where the electric field in SiPM cells could trigger the avalanche discharge by incident photons~\cite{HamamaMPPC}. And the over-voltage (V$_{over}$) could be defined by Eq.~\ref{Eq.Vover}:
\begin{equation}
V_{over}=V_{bias}-V_{bd}
\label{Eq.Vover}
\end{equation}

Where V${_{bias}}$ is the bias voltage of the SiPM array provided by the DC power supply. When the temperature is constant, the SiPM gain is positively related to V$_{over}$. So the gain would drop to zero when V${_{bias}}$=V${_{bd}}$. The most straightforward approach to estimate V${_{bd}}$ is calculating the X-intercept of the function of the SiPM gain and V${_{bias}}$ under a specific temperature. 

\begin{figure}[htbp]
	\centering
	\includegraphics[width=11cm]{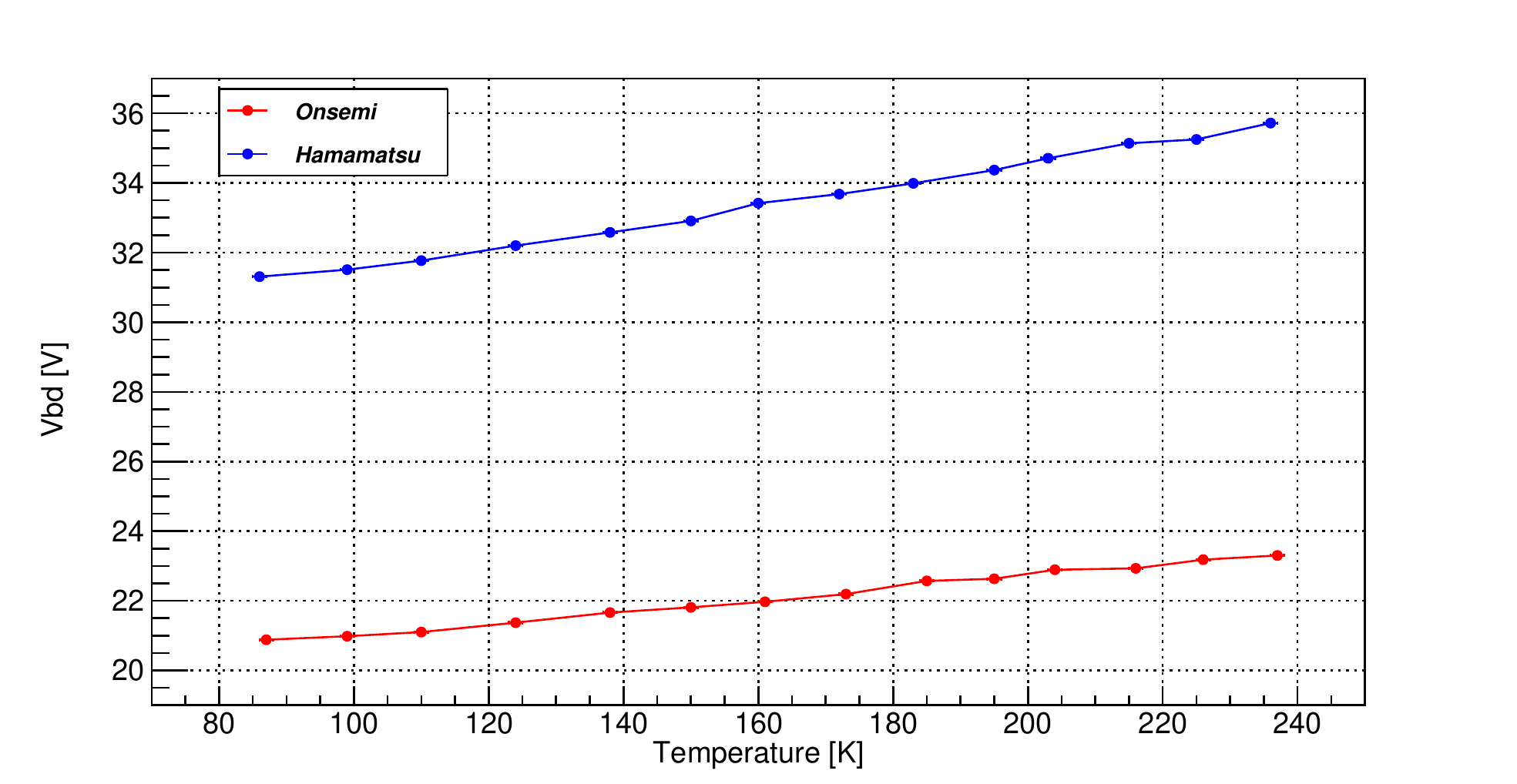}
	\caption{\label{Vbd} V${_{bd}}$ tendency with different temperatures from 237~K to 87~K. At 87~K, the V${_{bd}}$ of the Hamamatsu SiPM array is 31.3~V, and the Onsemi SiPM array is 20.88V. Errors in the X-axis are the accuracy of the temperature system, which is ±1~K. And errors in the Y-axis are statistical only but too small to see.} 	
\end{figure}

The tendency of V${_{bd}}$ with different temperatures has been plotted in Fig.~$\ref{Vbd}$. From 237~K to 87~K, the change rate of the Hamamatsu SiPM array and Onsemi SiPM array are about 0.3~V/10~K and 0.16~V/10~K, respectively. At the temperature of interest, 87~K, the V${_{bd}}$ of the Hamamatsu SiPM array is 31.3~V, and the Onsemi SiPM array is 20.8~V.

\begin{figure}[htbp]
	\centering
	\includegraphics[width=9.5cm]{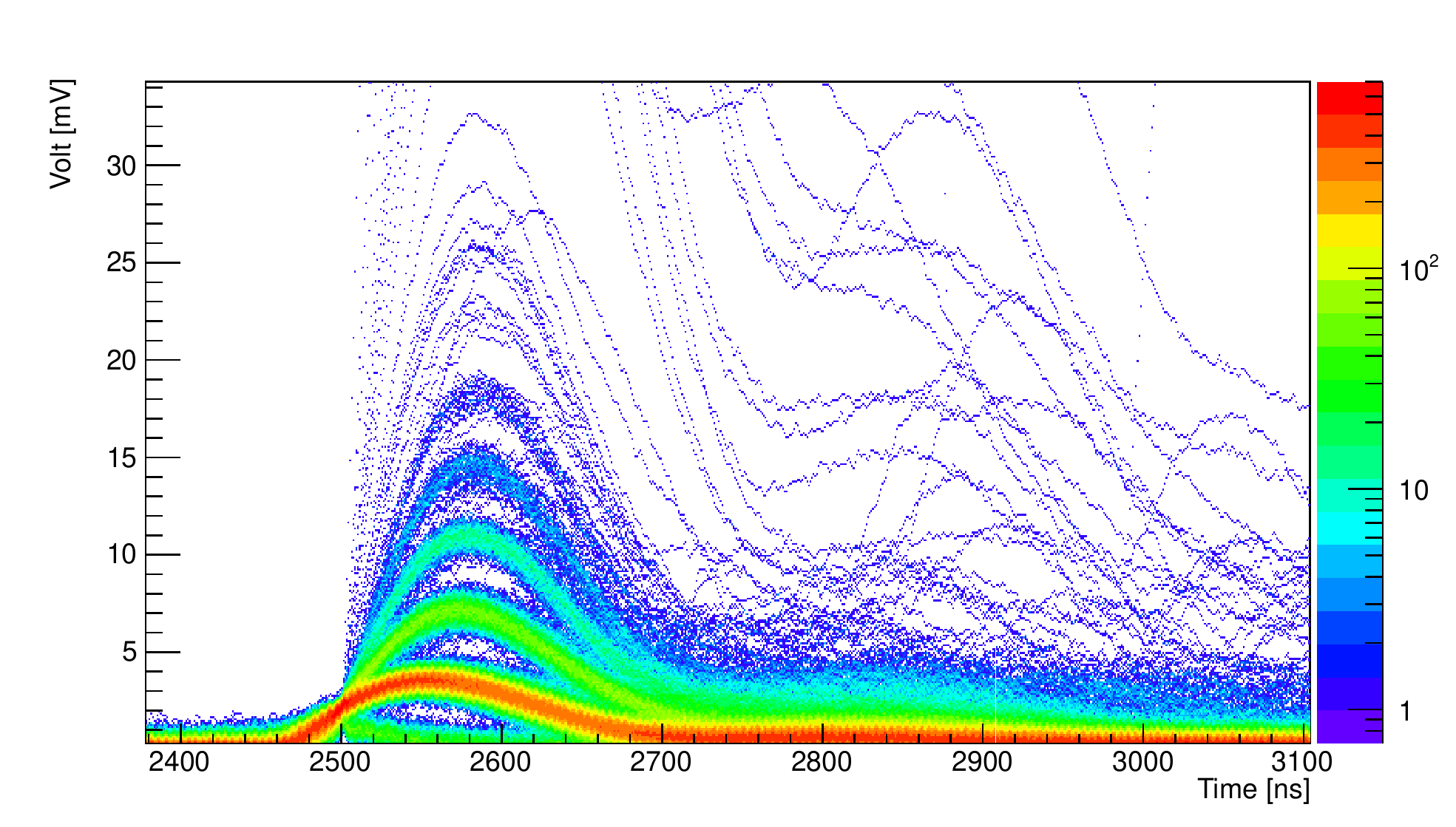}
	\qquad
	\includegraphics[width=9.5cm]{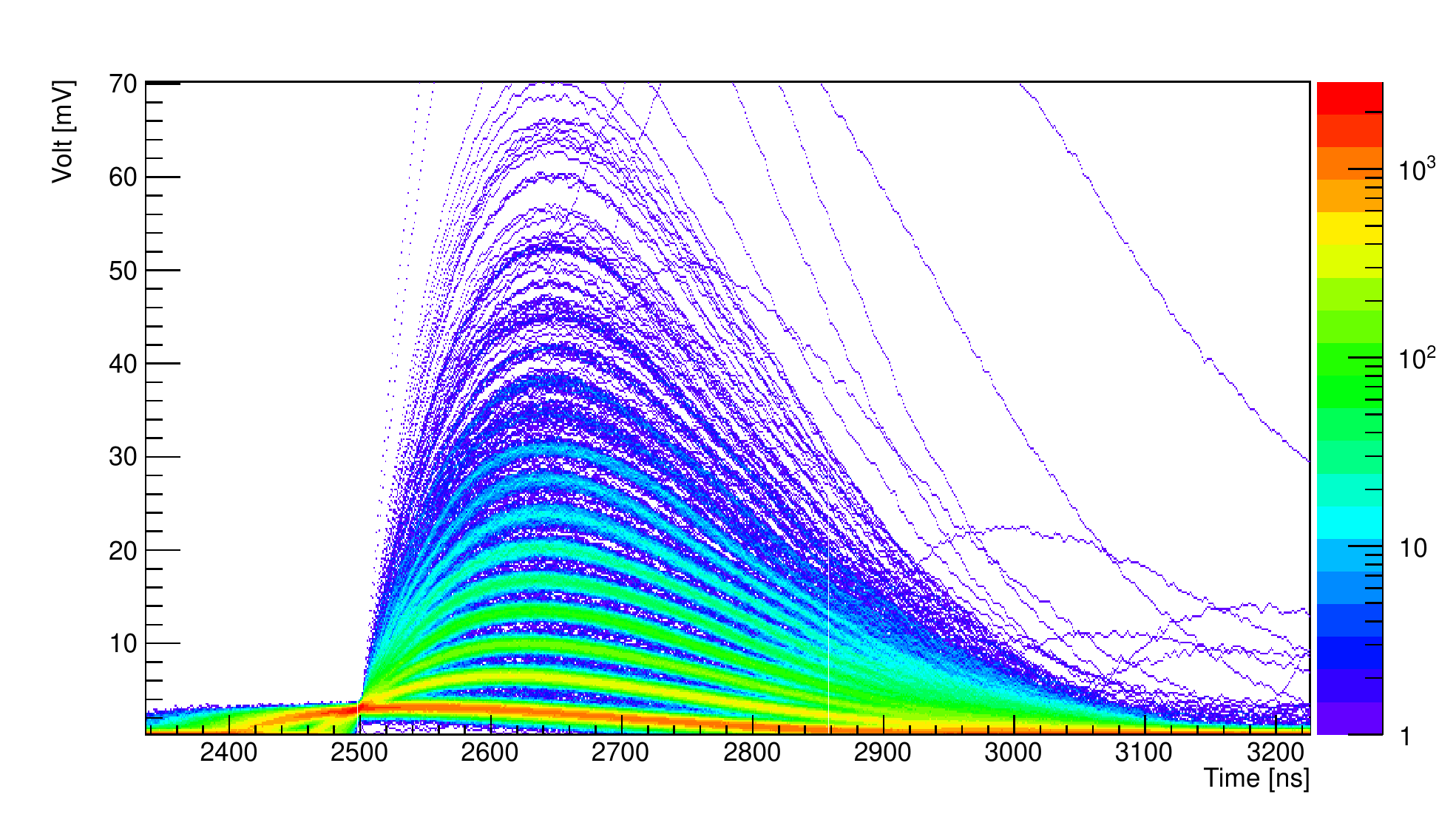}
	\caption{\label{array_SPE} Top: The overlap of five thousand dark signals of the 4 $\times$ 4 Hamamatsu S14161-6050HS SiPM array at liquid argon temperature. Bottom: twenty thousand dark signals of the 4 $\times$ 4 Onsemi J-Series 60035 SiPM array at liquid argon temperature. Excellent photoelectron consistencies were observed, especially the Onsemi SiPM array.}	
\end{figure}

\subsection{Performance of single photoelectron (SPE)}

The performances of SPE of two SiPM arrays were analyzed by recording dark signals in the SST at liquid argon temperature. Both SiPM arrays showed great photoelectron consistency, as expected. For instance, two graphs including numerous dark counting signals were plotted in Fig.~$\ref{array_SPE}$. The top graph in Fig.~$\ref{array_SPE}$ consists of about five thousand SPE pulses from Hamamatsu S14161-6050HS SiPM array. And the bottom graph is the profile composed of around twenty thousand SPE pulses from the Onsemi J-Series 60035 SiPM array. The great PE consistency makes the profile of multiple PE explicit, as shown in Fig.~$\ref{array_SPE}$, especially in the graph of the Onsemi SiPM array. 

\begin{figure}[htbp]
	\centering
	\includegraphics[width=6.5cm]{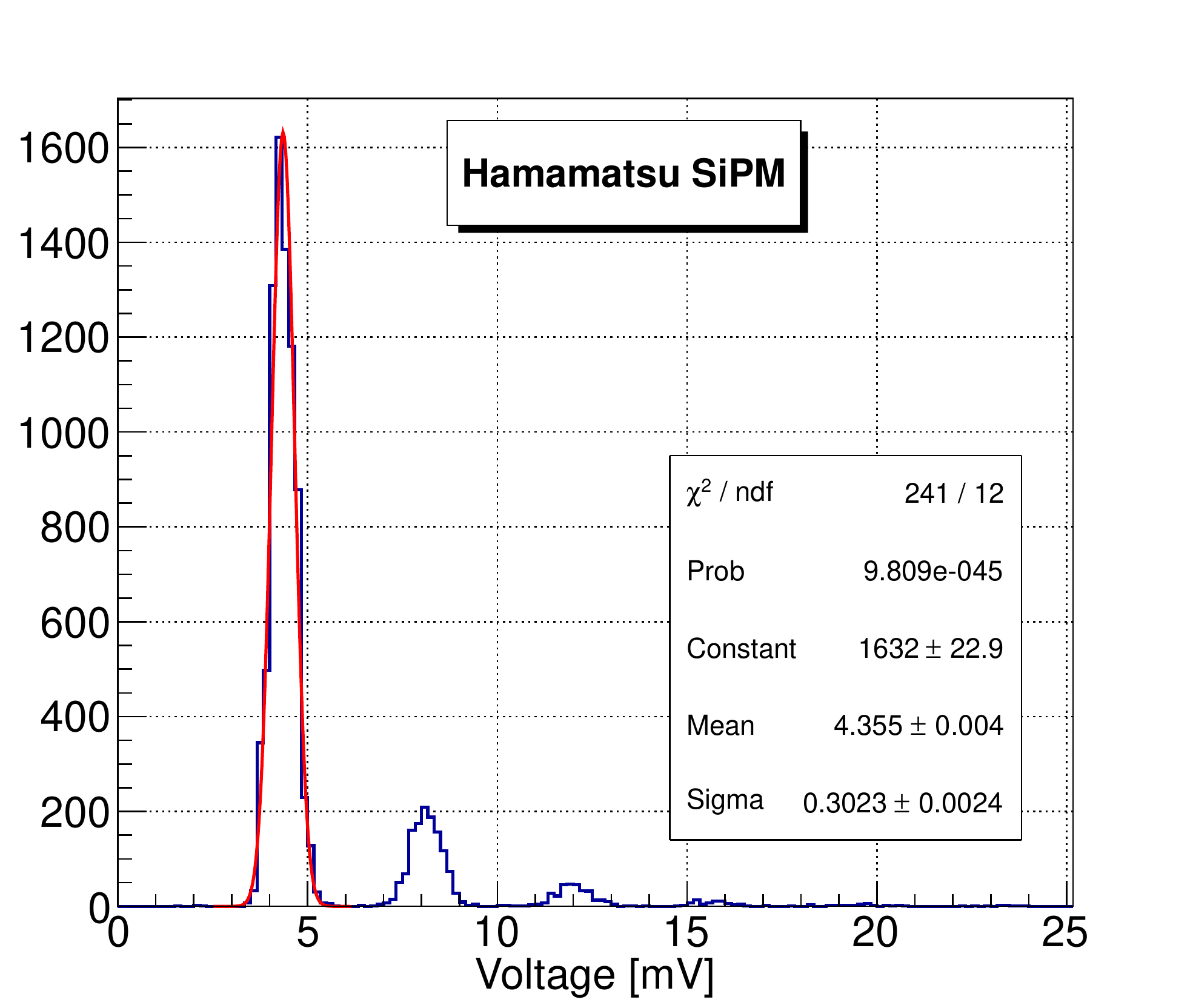}
	\includegraphics[width=6.5cm]{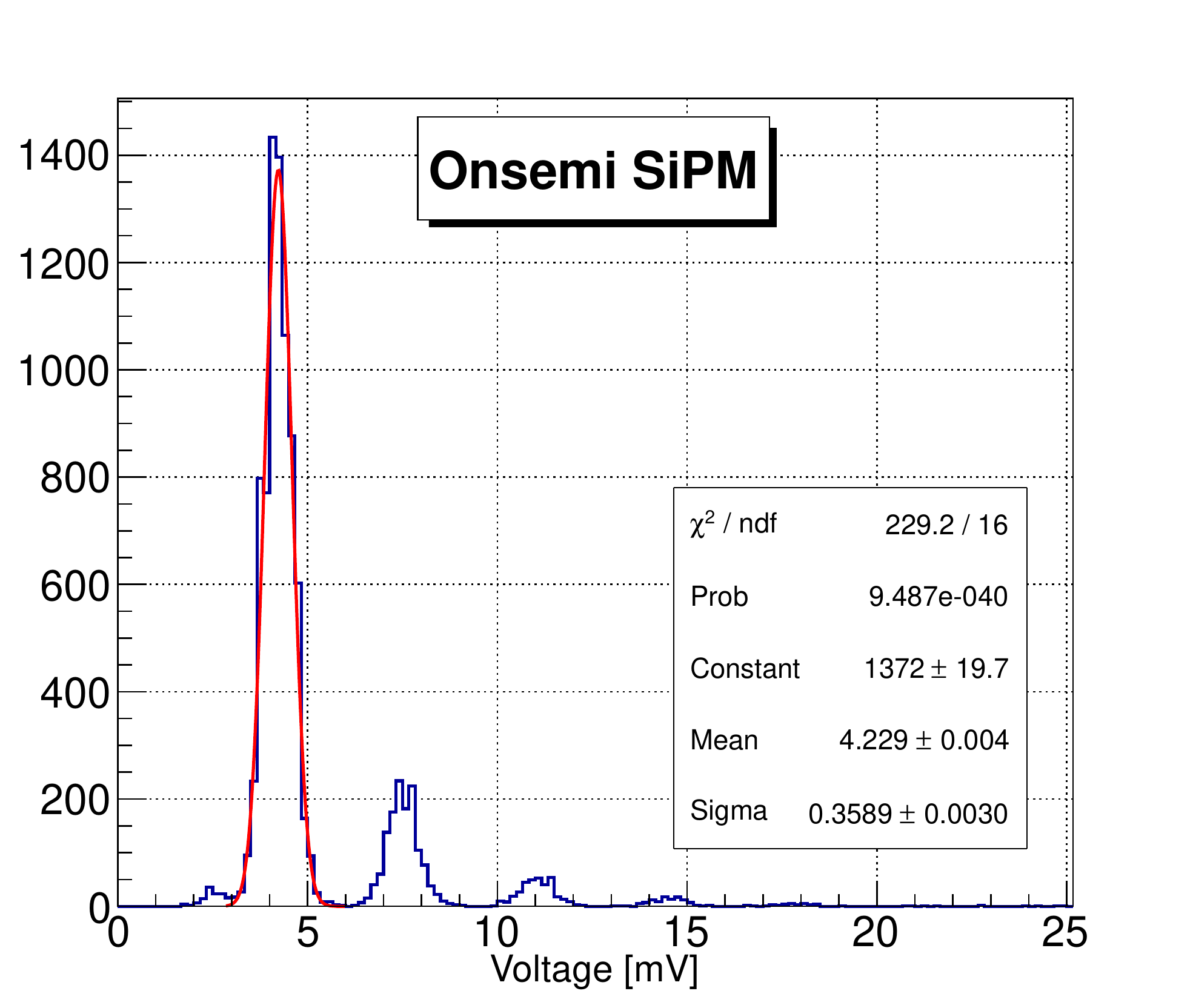}
	\caption{\label{array_Amplitud} SPE amplitude distributions of two SiPM arrays at 87K. The red curves are Gaussian fitting results for the SPE peak. They were measured with V${_{over}}$=4V. }	
\end{figure}

The rise and fall times of PE waveforms can be calculated as follows:
\begin{equation}
t_{rise}=R_{s}\times C_{d}
\label{Eq.time}
\end{equation}
\begin{equation}
t_{fall}=R_{q}\times C_{d}
\label{Eq.time}
\end{equation}

Where R${_{s}}$ is the series resistor, R${_{q}}$ is the quenching resistor, and C${_{d}}$ is the capacitance of Geiger-mode SiPM. Fig.~$\ref{array_SPE}$ indicates a larger C${_{d}}$ of the Onsemi J-Series 60035 SiPM array because of the longer rise and fall times of its SPE. They are about 150~ns and 500~ns separately, compared with about 80~ns and 150~ns for the Hamamatsu S14161-6050HS SiPM array. Thus, for achieving a similar amplitude of SPE, G${_{amp}}$ of the Onsemi J-Series 60035 SiPM array had to be adjusted. R${_F}$ of two SiPM arrays are different, and they were set to 2~k~$\Omega$ and 3.9~k~$\Omega$ for Hamamatsu and Onsemi, respectively. Fig.~$\ref{array_Amplitud}$ shows the amplitude distributions of two SiPM arrays with V${_{over}}$=4~V at 87~K. Similar amplitudes of SPE were achieved by adjusting G${_{amp}}$. 

\begin{figure}[htbp]
	\centering
	\includegraphics[width=9cm]{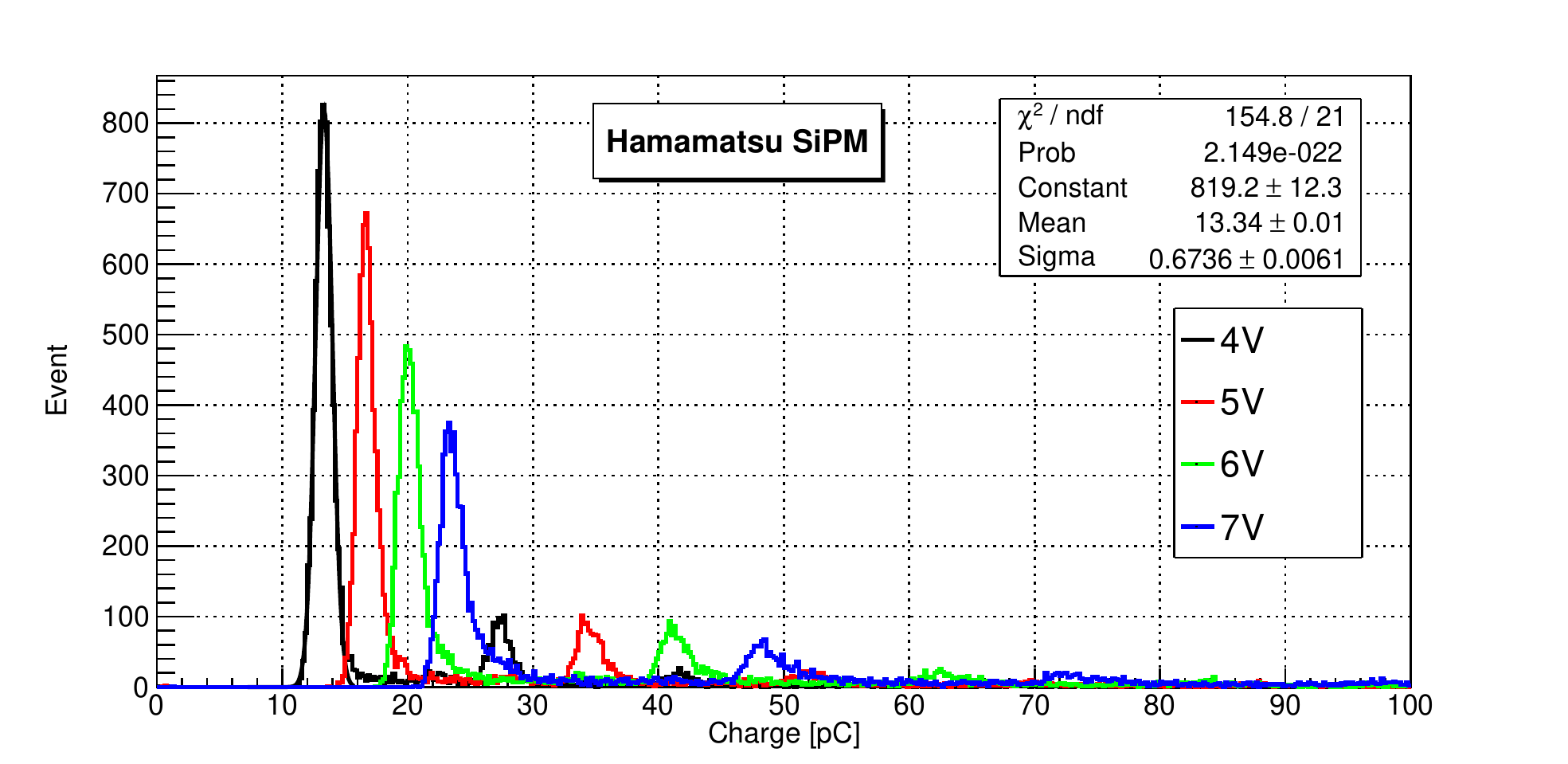}
	\qquad
	\includegraphics[width=9cm]{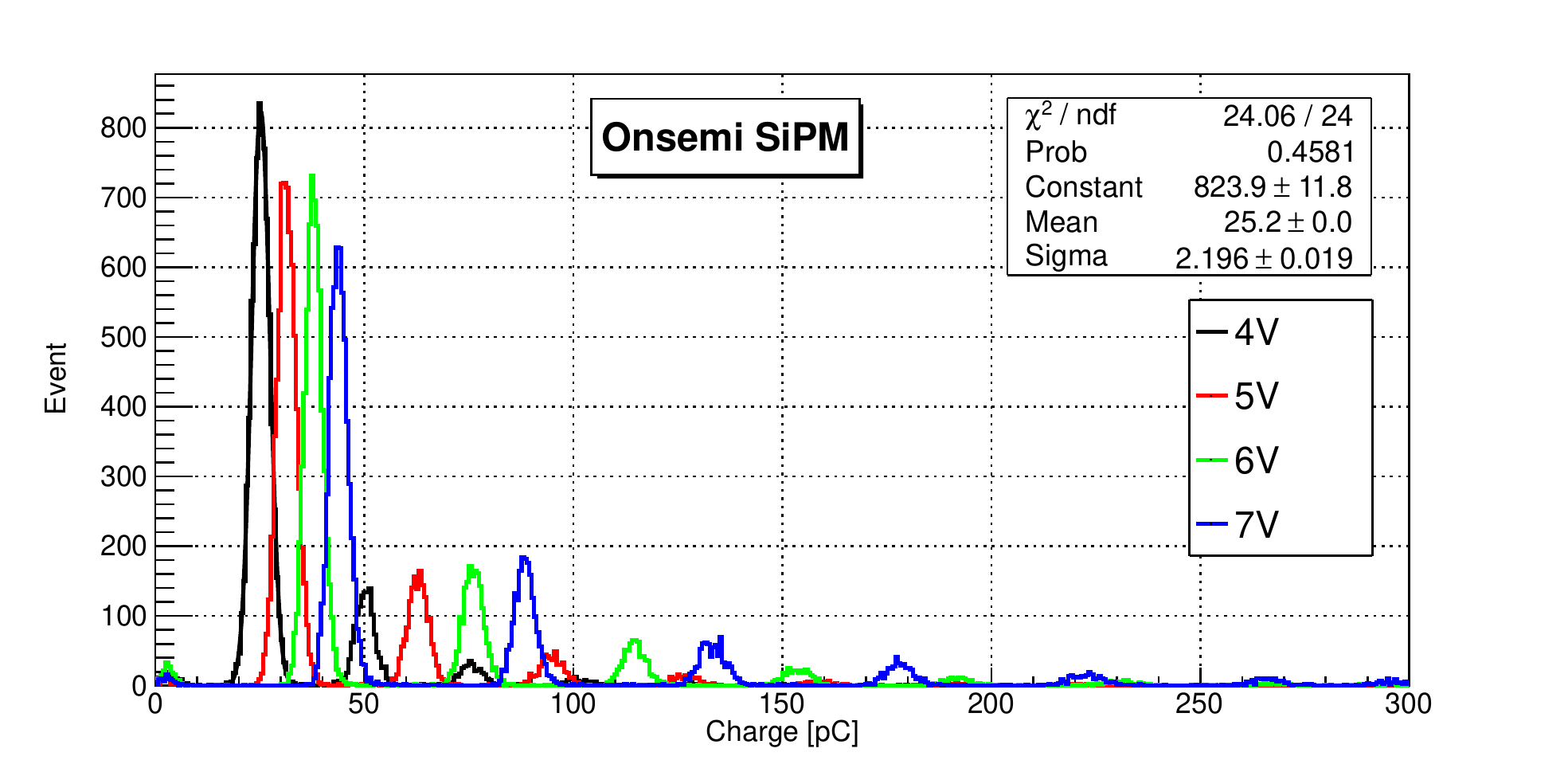}
	\caption{\label{array_Spectrum} Photoelectron energy spectra with different V${_{over}}$ filled by ten thousand dark signals at 87K. The four highest peaks with different colors left side of the graphs are the energy distributions of single PE from two types of SiPM arrays. And the data upper right of the graphs is the Gaussian fit of single PE peaks at V${_{over}}$ = 4~V.}	
\end{figure}

Furthermore, the SPE energy spectra of two SiPM arrays were plotted in Fig.~$\ref{array_Spectrum}$. Four colors represent the spectra with different V${_{over}}$. A character could be observed that the G$_{SiPM}$ is positively related to V${_{over}}$. Another characteristic in Fig.~$\ref{array_Spectrum}$ is the count decreasing of the single photoelectron peaks with the increase of V${_{over}}$. This phenomenon attributes to the rise of correlated signals. The probability of double or more discharges by just one incident photon would get more significant with high V${_{over}}$.

\subsection{Dark count rate (DCR)}

\begin{figure}[htbp]
	\centering
	\includegraphics[width=9cm]{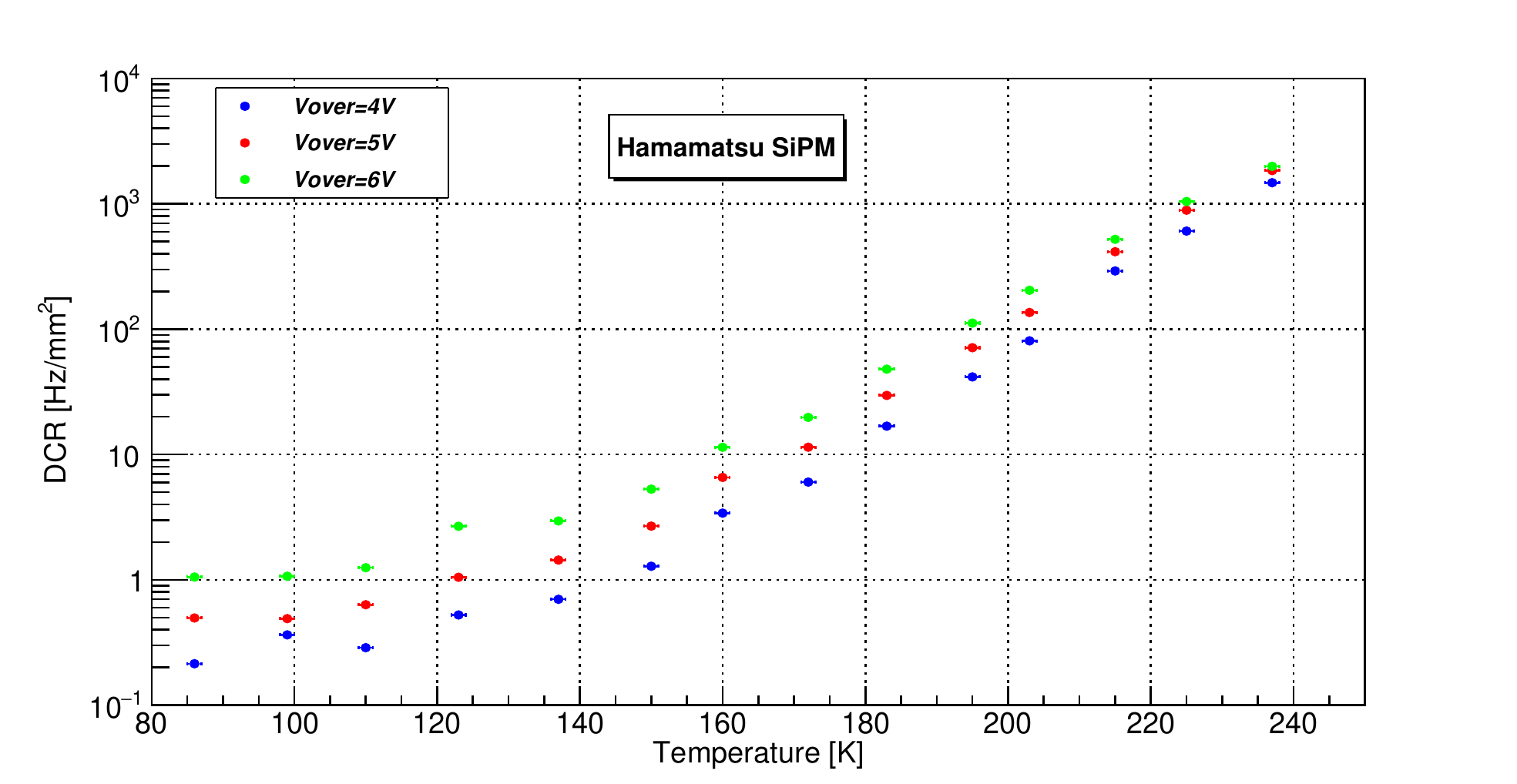}
	\qquad
	\includegraphics[width=9cm]{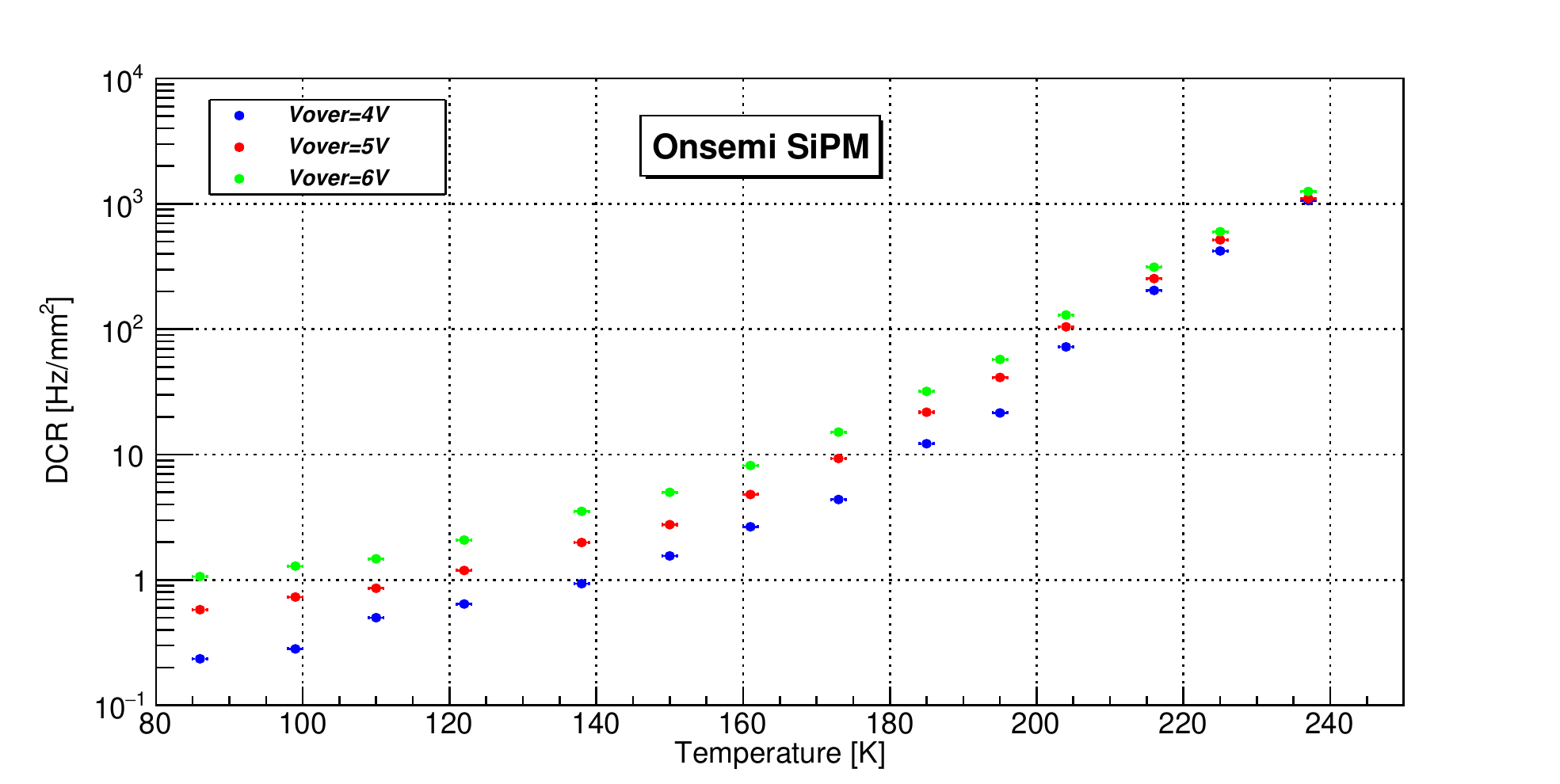}
	\caption{\label{array_DCR} Distribution of DCR at different temperatures when V${_{over}}$ = 4~V, V${_{over}}$ = 5~V, and V${_{over}}$ = 6~V. Up: Changes of the Hamamatsu SiPM array DCR. The errors in the X-axis are the accuracy of the temperature system, which is ±1 K, the errors in the Y-axis are statistical only. Down: Changes of the Onsemi SiPM array DCR. Errors of the X-axis are the accuracy of the temperature system, which is ±1 K. Errors of the Y-axis are statistical only.}	
\end{figure}

The prominent noise in the SiPM array is the dark count rate (DCR), primarily due to thermal electrons generated in the active volume. The DCR is a function of the active area, V${_{over}}$ and temperature~\cite{OnsemiSiPM}.
Fig.~$\ref{array_DCR}$ shows the change of DCR with the temperature at different V${_{over}}$. When the temperature is constant, the DCR decreases with the decreases of the V${_{over}}$. 

The upper graph in Fig.~$\ref{array_DCR}$ shows that the DCR of the Hamamatsu SiPM array at 87K is $\sim$0.214~Hz/mm$^{2}$, $\sim$0.496~Hz/mm$^{2}$ and $\sim$1.053~Hz/mm$^{2}$ when V${_{over}}$ is 4~V, 5~V and 6~V, respectively. On contrast, the DCR of the Onsemi SiPM array at 87K is $\sim$0.235~Hz/mm$^{2}$, $\sim$0.579~Hz/mm$^{2}$ and $\sim$1.065~Hz/mm$^{2}$ when V${_{over}}$ is 4~V, 5~V and 6~V, respectively. There is little difference between the two types of SiPM arrays, but they are in the same order.

\subsection{Signal to noise ratio (SNR)}

\begin{figure}[htbp]
	\centering
	\includegraphics[width=11cm]{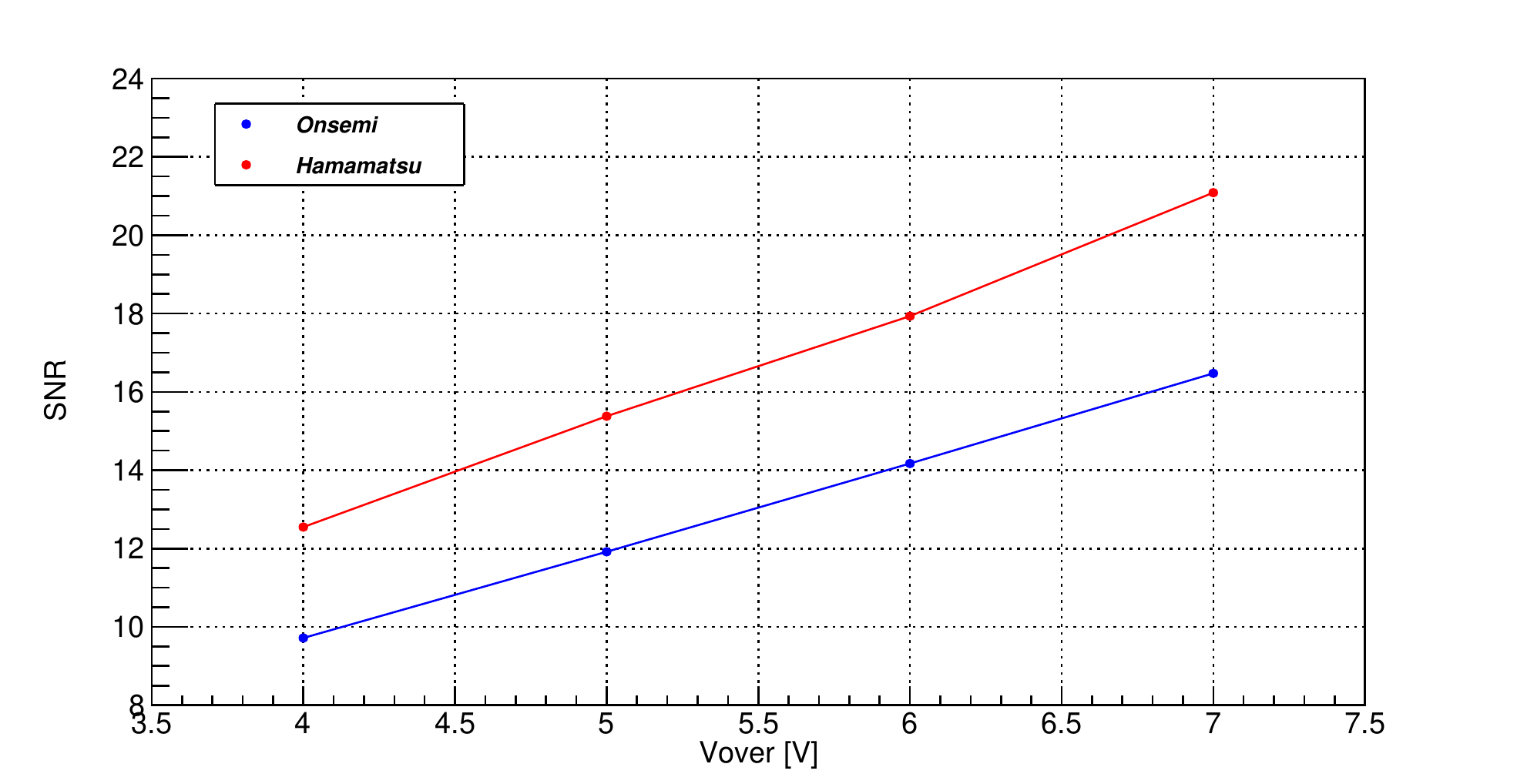}
	\caption{\label{SNR} SNR distribution for two types of SiPM arrays with different V${_{over}}$ at 87~K. The temperature changes cause the errors on the X-axis, and the errors on the Y-axis are statistical only.} 	
\end{figure}

SNR is defined as the ratio of the single photoelectron amplitude to the standard deviation of the noise, as shown in Eq.\ref{Eq.SNR}. 

\begin{equation}
	SNR=\frac{V_{SPE}}{\sigma_{baseline}}
	\label{Eq.SNR}
\end{equation}
Where V${_{SPE}}$ is the SPE amplitude mentioned in Fig.~$\ref{array_Amplitud}$. Fig.~$\ref{SNR}$ shows the change of SNR with different V${_{over}}$. It is expected that the SNR is positively related to V${_{over}}$ because the SPE amplitude increase with  V${_{over}}$. And $\sigma_{baseline}$ is independent of it. As a result, the SNR from the Hamamatsu SiPM array is slightly better than from the Onsemi SiPM array.

\subsection{Estimatation of correlated signals}
Correlated signals like afterpulsing (AP) and direct cross-talk (DiCT) are other components that could affect the performance of SiPM arrays negatively, which is related to V${_{over}}$. It was introduced in Ref~\cite{SiPM-applications,VUV4_SiPM,OnsemiSiPM,Study_AP} in detail. The high probability of AP and DiCT would worsen the energy resolution of SiPM output signals and cause an over-estimation of the light yield of scintillators.

\subsubsection{Afterpulsing}

\begin{figure}[htbp]
	\centering
	\includegraphics[width=13cm]{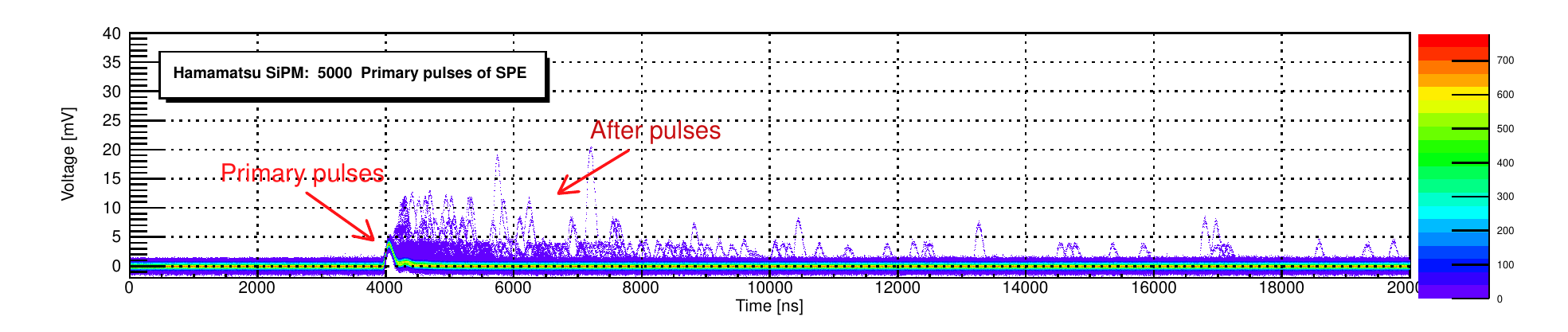}
	\caption{\label{Hama_CT} AP performance of the Hamamatsu SiPM array at 87~K, by overlapping 5000 SPE signals in the 20~$\mu$s window, with V${_{over}}$ = 4~V. The primary pulses and AP positions were marked out by two red arrows separately.} 	
\end{figure}

\begin{figure}[htbp]
	\centering
	\includegraphics[width=11cm]{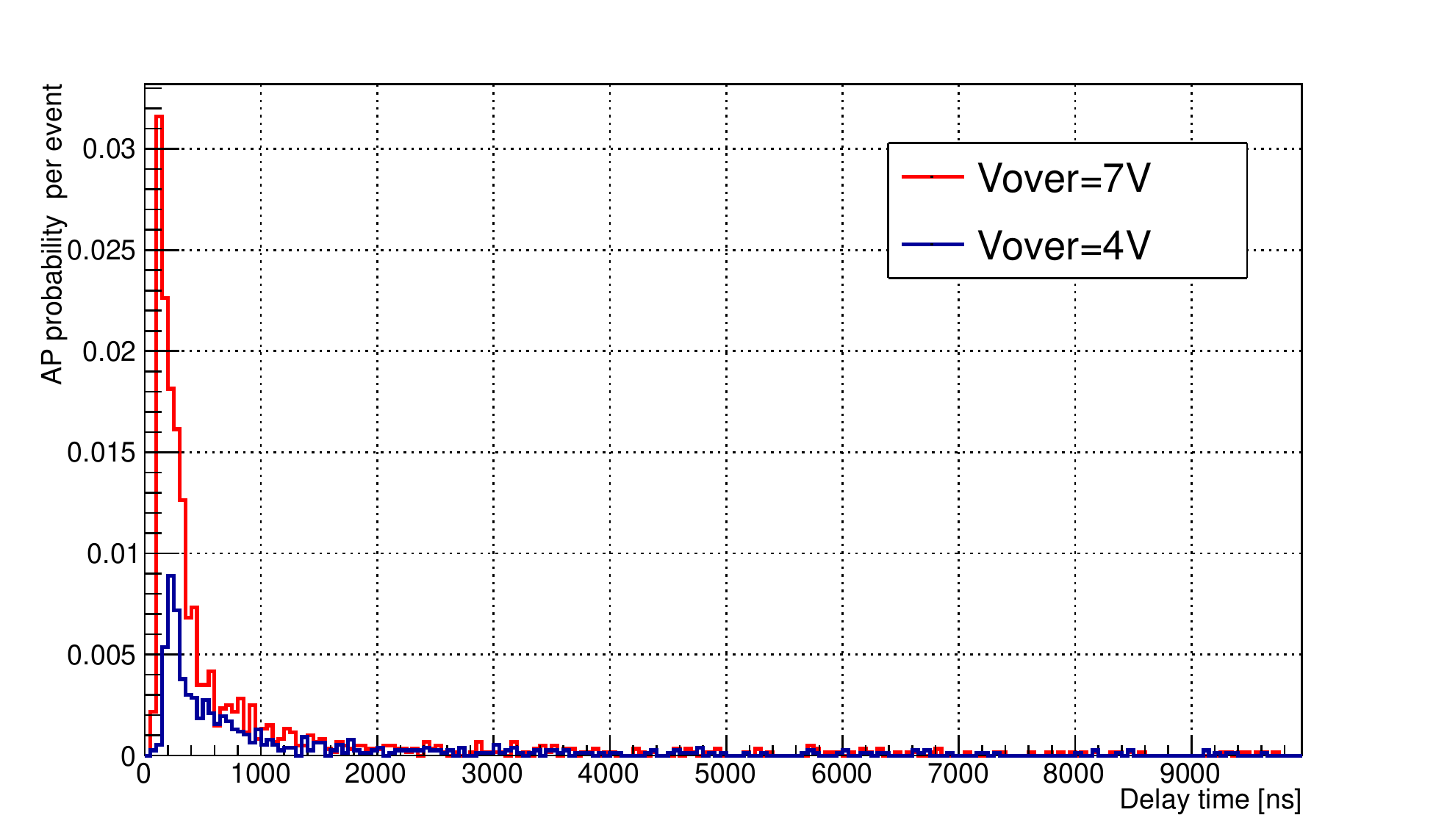}
	\caption{\label{Hama_AP_delaytime} Distribution of the AP probability of the Hamamatsu SiPM array with the delay time at 87K. } 	
\end{figure}

AP is caused by the release of the trapped electrons during discharge. Fig~$\ref{Hama_CT}$ is the example of AP performance of the Hamamatsu SiPM array by overlapping five thousand SPE pulses. In the graph, primary SPE pulses' arrival time was set to 4000~ns. A lot of AP could be observed behind primary pulses. They both were marked out by two red arrows separately. A prominent characteristic of AP shown in Fig~$\ref{Hama_CT}$ is that the amount of AP is inversely proportional to the delay time. Fig.~$\ref{Hama_AP_delaytime}$ demonstrated the distribution of AP probability and delay time between primary SPE pulses and AP. The figure also shows that the probability of AP is proportional to the value of V${_{over}}$. Thus, the bias voltage of SiPM arrays should be appropriately set to control the ratio of AP.

\begin{figure}[htbp]
	\centering
	\includegraphics[width=13cm]{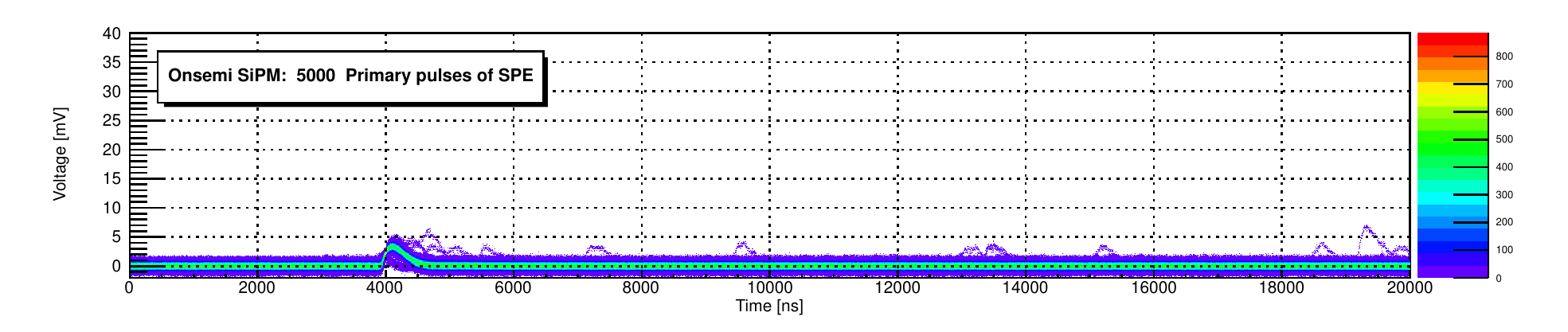}
	\caption{\label{Sensl_CT} AP performance of Onsemi SiPM array at 87k, V${_{bias}}$=24.8V.} 	
\end{figure}
\begin{figure}[htbp]
	\centering
	\includegraphics[width=11cm]{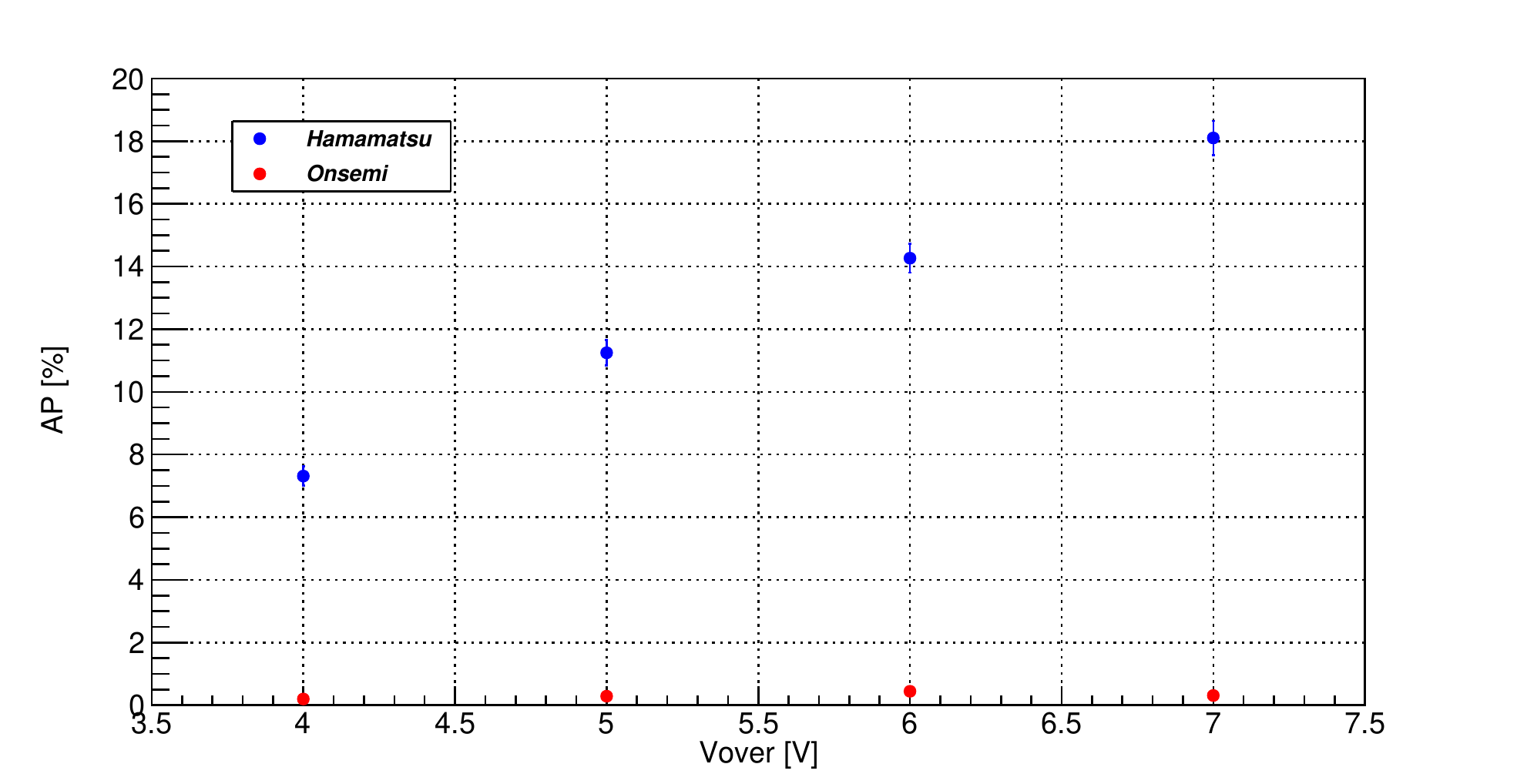}
	\caption{\label{VoverVSAP} The probabilities of AP tendency with V${_{over}}$ at 87~K. The temperature variation causes the errors on the X-axis, and the errors on the Y-axis are statistical only.} 	
\end{figure}

In contrast, another example of AP performance is from the Onsemi SiPM array and includes five thousand SPE pulses as well in Fig~$\ref{Sensl_CT}$. The amount of AP from the Onsemi SiPM array is significantly less than from the Hamamatsu SiPM array. In Fig~$\ref{Sensl_CT}$, there are only about fifteen AP behind a total of five thousand primary pulses at V${_{over}}$ = 4~V. Conclusively, Fig~$\ref{VoverVSAP}$ shows the change of the AP probability with different V${_{over}}$ of two types of SiPM arrays. They both increase with the increase of V${_{over}}$, while the AP probability from the Onsemi SiPM array changes slightly.

\subsubsection{Direct crosstalk}

Sometimes, multiple avalanche diodes operated in Geiger mode are excited spontaneously by one photon, defined as Direct Crosstalk (DiCT). It would cause the overestimation of particle energies as well. The probability of DiCT could be estimated by:  
\begin{equation}
	P(DiCT)=\frac{N(1.5PE)}{N(0.5PE)}
	\label{Eq.DiCT}
\end{equation}

\begin{figure}[htbp]
	\centering
	\includegraphics[width=11cm]{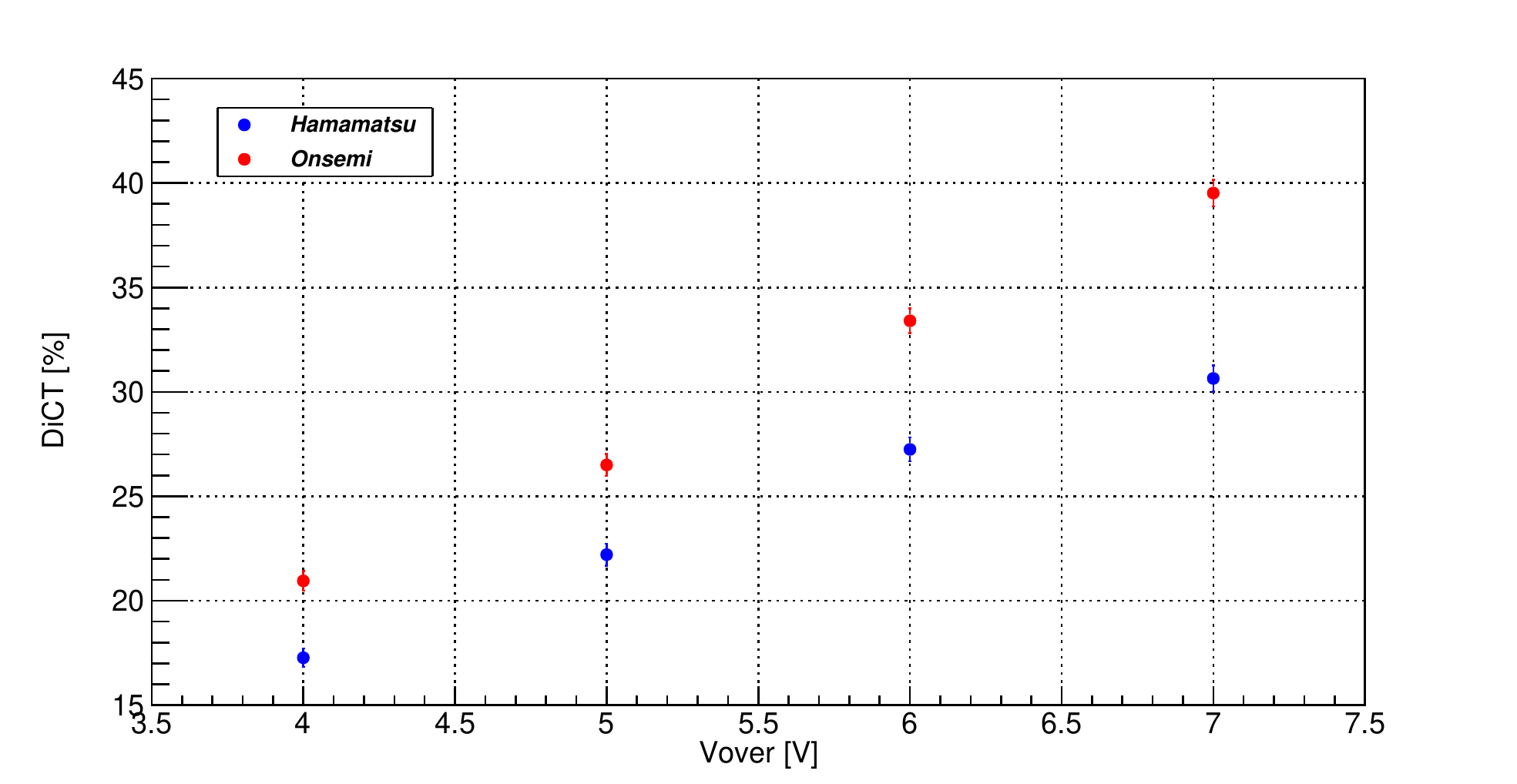}
	\caption{\label{VoverVSDiCT} DiCT probability of two types of SiPM arrays versus V${_{over}}$. The temperature variation causes errors in the X-axis, and errors in the Y-axis are statistical only.} 	
\end{figure}

The changes of the DiCT probability with different V${_{over}}$ are shown in Fig~$\ref{VoverVSDiCT}$. As the performance of AP, the probability of DiCT is linearly related to V${_{over}}$. The DiCT probability of the Onsemi SiPM array is higher than the Hamamatsu SiPM array.

In conclusion, in the aspect of correlated signal performances, two SiPM arrays have their advantages and disadvantages. The Hamamatsu SiPM array has a higher AP probability, and the Onsemi SiPM array gets a higher DiCT probability. Moreover, they all suffer from the increase of V${_{over}}$.

\section{Conclusion}

In this paper, we successfully achieved the operation of SiPM arrays and their cryogenic electronics at liquid argon temperature. The power dissipation of the amplifiers is less than 10~$\mu$W/mm$^2$. Two SiPM arrays from different vendors worked steadily and showed excellent SPE performances. At liquid argon temperature, the DCR from both SiPM is $\sim$ 0.2 Hz/mm$^2$ at V${_{over}}$ = 4~V. Moreover, their SNRs are both above 10. 

The Hamamatsu SiPM array performs better at the two characteristics above slightly. While the Onsemi SiPM array gets a lower probability of correlated signals. Primarily, it achieved nearly zero AP probability when working at liquid argon temperature, though the probability of DiCT is higher than the Hamamatsu SiPM array.

\section{Acknowledgments}

This work is supported by the National Key R \& D Program of China (Grant No. 2016YFA0400304) and the National Natural Science Foundation of China (Grant No. 12020101004, Grant No.11975257, and Grant No.12175247).



\begin{thebibliography}{00}
\bibitem{SiPM-his}
Renker.~Dieter, Geiger-mode avalanche photodiodes, history, properties and problems, Nucl.~Instrum.~Meth.~A 567 (2006) 48-56.


\bibitem{C-DUNE}
Minotti.~Alessandro and DUNE collaboration and others, Characterization of the {DUNE} photodetectors and study of the event burst phenomenon, Journal of Physics: Conference Series 2156 (2021) 012242.

\bibitem{SiPM-DUNE}
Falcone.~A, Andreani.~A, Bertolucci.~S, Brizzolari.~C, Buckanam.~N, Capasso.~M, Cattadori.~C, Carniti.~P, Citterio.~M, Francis.~K, others, Cryogenic SiPM arrays for the DUNE photon detection system, Nuclear Instruments and Methods in Physics Research Section A: Accelerators, Spectrometers, Detectors and Associated Equipment 985 (2021) 164648.


\bibitem{DarkSide-20k}
Bottino~Bianca, DarkSide-20k and the Future Liquid Argon Dark Matter Program, PoS, EPS-HEP2021 169 (2022).


\bibitem{SiPM-Japan}
Aoyama.~Kazutaka, Tanaka.~Masashi, Kimura.~Masato, Yorita.~Kohei, Development of a liquid argon detector with high light collection efficiency using tetraphenyl butadiene and a silicon photomultiplier array, Progress of Theoretical and Experimental Physics 2022 (2022) 043H01.

\bibitem{SiPM-applications}
Buzhan.~P, Dolgoshein.~B, Filatov.~L, Ilyin.~A, Kaplin.~V, Karakash.~A, Klemin.~S, Mirzoyan.~R, Otte.~AN, Popova.~E and others, Large area silicon photomultipliers: Performance and applications, Nuclear Instruments and Methods in Physics Research Section A: Accelerators, Spectrometers, Detectors and Associated Equipment 567 (2006) 78-82.

\bibitem{OnsemiSiPM}
{An~Introduction~to~the~Silicon~photomultiplier}, https://www.onsemi.com.

\bibitem{VUV4_SiPM} L.~Wang, M.Y.~Guan, H.J.~Qin, C.~Guo, X.L.~Sun, C.G.~Yang, Q.~Zhao, J.C.~Liu, P.~Zhang, Y.P.~Zhang, W.X.~Xiong, Y.T.~Wei, Y.Y.~Gan, J.J.~Li, Characterization of VUV4 SiPM for liquid argon detector,  Journal of instrumentation 16 (2021) P07021.

\bibitem{HamaDS}
{Hamamatsu S14161 SiPM datasheet}, https://www.hamamatsu.com.


\bibitem{OnsemiDS}
{Onsemi J-Series SiPM datasheet}, https://www.onsemi.com.


\bibitem{LMH6629DS}
{LMH6629 datasheet}, https://www.ti.com.

\bibitem{TIA_con}
Ramus.~Xavier, Transimpedance considerations for high-speed amplifiers, Application Report SBOA122. Texas Instruments (2009) 97.

\bibitem{SiPMLAr}
Guo.~Cong, Guan.~MY, Sun.~XL, Xiong.~WX, Zhang.~Peng, Yang.~CG, Wei.~YT, Gan.~YY, Zhao.~Q, The liquid argon detector and measurement of SiPM array at liquid argon temperature, Nuclear Instruments and Methods in Physics Research Section A: Accelerators, Spectrometers, Detectors and Associated Equipment 980 (2020) 164488.


\bibitem{Preamplifier}
D’Incecco.~Marco, Galbiati.~Cristiano, Giovanetti.~Graham K, Korga.~George, Li.~Xinran, Mandarano.~ Andrea, Razeto.~Alessandro, Sablone.~Davide, Savarese.~Claudio, Development of a very low-noise cryogenic preamplifier for large-area SiPM devices, IEEE Transactions on Nuclear Science 65 (2018) 1005-1011.

\bibitem{HamamaMPPC}
{Hamamatsu MPPC technical note}, https://www.hamamatsu.com.

\bibitem{Study_AP}
Otono.~Hidetoshi, Study of MPPC at liquid nitrogen temperature, PoS (2007) 007.



\end{thebibliography}



\end{document}